\documentclass[a4paper]{article}
\usepackage{graphicx}
\usepackage{amsmath}

\begin{document}

\title{The characteristic function and entanglement of optical evolution}
\author{Xiao-yu Chen \\
{\small {Lab. of Quantum Information, China Institute of Metrology,
Hangzhou, 310034, China}}}
\date{}
\maketitle

\begin{abstract}
The master equation of quantum optical density operator is transformed to
the equation of characteristic function. The parametric amplification and
amplitude damping as well as the phase damping are considered. The solution
for the most general initial quantum state is obtained for parametric
amplification and amplitude damping. The purity of one mode Gaussian system
and the entanglement of two mode Gaussian system are studied.


\end{abstract}

\section{Introduction}

Quantum information with continuous variables (CV) \cite{Braunstein1} \cite
{Braunstein2} is a flourishing field, as shown by the spectacular
implementations of deterministic teleportation schemes\cite{Vaidman} \cite
{Furusawa} \cite{Takei} \cite{vanLoock1}, quantum key distribution protocols
\cite{Grosshans}, entanglement swapping \cite{Takei} \cite{vanLoock2}, dense
coding \cite{Ban}, quantum state storage \cite{Julsgaard} and quantum
computation \cite{Lloyd} processes in quantum optical settings . The crucial
resource enabling a better-than-classical manipulation and processing of
information is CV entanglement. In all such practical instances the
information and entanglement contained in a given quantum state of the
system, so precious for the realization of any specific task, is constantly
threatened by the unavoidable interaction with the environment. Such an
interaction entangles the system of interest with the environment, causing
some amount of information to be scattered and lost in the environment. The
overall process, corresponding to a non unitary evolution of the system, is
commonly referred to as decoherence \cite{Furusawa} \cite{Takei}. In this
work we study the decoherences of general states of continuous variable
systems whose evolutions are ruled by optical master equations. We will
consider the parametric amplification, amplitude damping and phase damping
\cite{Kinsler} of a general quantum CV state in the fashion of quantum
characteristic function. The main starting point for this research work was
the result of Lindblad's bounded generator of a completely positive quantum
dynamical semigroup \cite{Lindblad}. Quantum characteristic function had
been used to treat the amplitude damping of one mode squeezed states \cite
{Marian}.

\section{Time evolution of characteristic function}

The density matrix obeys the following master equation \cite{Kinsler} \cite
{Lindblad}\cite{Walls}
\begin{equation}
\frac{d\rho }{dt}=-\frac i\hbar [H,\rho ]+(\mathcal{L}_1\mathcal{+L}_2)\rho .
\end{equation}
with the quadratic Hamiltonian
\begin{equation}
H=\hbar \sum_{jk}\frac i2(\eta _{jk}a_j^{\dagger }a_k^{\dagger }-\eta
_{jk}^{*}a_ja_k)
\end{equation}
where $\eta $ is a complex symmetric matrix. In the single-mode case, this
Hamiltonian describes two-photon downconversion from an undepleted
(classical) pump\cite{Walls}. The full multi-mode model describes
quasi-particle excitation in a BEC within the Bogoliubov approximation \cite
{Leggett}. This item represents the parametric amplifier. While the
amplitude damping is described by $\mathcal{L}_1,$

\begin{eqnarray}
\mathcal{L}_1\rho &=&\sum_j\frac{\Gamma _j}2\{(\overline{n}_j+1)L[a_j]\rho +%
\overline{n}_jL[a_j^{\dagger }]\rho \\
&&-w_j^{*}M[a_j]\rho -w_jM[a_j^{\dagger }]\rho \}.  \nonumber
\end{eqnarray}
where the Lindblad super-operators are defined as $L[\widehat{o}]\rho \equiv
$ $2\widehat{o}\rho \widehat{o}^{\dagger }-\widehat{o}^{\dagger }\widehat{o}%
\rho -\rho \widehat{o}^{\dagger }\widehat{o}$ and $M[\widehat{o}]\rho \equiv
2\widehat{o}\rho \widehat{o}-\widehat{o}\widehat{o}\rho -\rho \widehat{o}%
\widehat{o}.$ The requirement of positivity of the density operator imposes
the constraint $\left| w_j\right| ^2\leq \overline{n}_j(\overline{n}_j+1).$
At thermal equilibrium, \textit{i.e. }$w_j=0,$ $\overline{n}_j$ is equal to
the average thermal photon number of the environment. If $w_j\neq 0,$ then
the bath $j$ is said to be `squeezed'\cite{Serafini}.

$\mathcal{L}_2$ describes the phase damping,
\begin{equation}
\mathcal{L}_2\rho =\sum_j\frac{\gamma _j}2L[a_j^{\dagger }a_j]\rho .
\end{equation}

We now transform the density operator master equation to the diffusion
equation of the characteristic function. We use characteristic function
because it is more convinent for our problem. One may use Glauber's
P-representation instead, but Glauber's P-representation does not exist for
some of the states (e.g.\cite{Marian} \cite{Holevo}), although the
generalised positive P-representation does always exist \cite{Drummond}. Any
quantum state can be equivalently specified by its characteristic function.
Every operator $\mathcal{A}\in \mathcal{B(H)}$ is completely determined by
its characteristic function $\chi _{\mathcal{A}}:=tr[\mathcal{AD}(\mu )]$
\cite{Petz}, where $\mathcal{D}(\mu )=\exp (\mu a^{\dagger }-\mu ^{*}a)$ is
the displacement operator, with $\mu =[\mu _1,\mu _2,\cdots ,\mu _s]$ $%
,a=[a_1,a_2,\cdots ,a_s]^T$ and the total number of modes is $s.$ It follows
that $\mathcal{A}$ may be written in terms of $\chi _{\mathcal{A}}$ as \cite
{Perelomov}: $\mathcal{A}=\int [\prod_i\frac{d^2\mu _i}\pi ]\chi _{\mathcal{A%
}}(\mu )\mathcal{D}(-\mu ).$ The density matrix $\rho $ can be expressed
with its characteristic function $\chi $. $\chi =tr[\rho \mathcal{D}(\mu )]$
. Multiplying $\mathcal{D}(\mu )$ to the master equation then taking trace,
the master equation of density operator will be transformed to the diffusion
eqation of the characteristic function. It should be noted that the complex
parameters $\mu _j$ are not a function of time, thus $\frac{\partial \chi }{%
\partial t}=tr[\frac{\partial \rho }{\partial t}\mathcal{D}(\mu )],$ the
parametric amplification part in the form of characteristic function will be
\cite{Walls}

\begin{equation}
\frac 12tr\{\sum_{jk}[\eta _{jk}a_j^{\dagger }a_k^{\dagger }-\eta
_{jk}^{*}a_ja_k,\rho ]D(\mu )\}=-\sum_{jk}(\eta _{jk}\mu _j^{*}\frac{%
\partial \chi }{\partial \mu _k}+\eta _{jk}^{*}\mu _j\frac{\partial \chi }{%
\partial \mu _k^{*}}).
\end{equation}

The master equation can be transformed to the diffusion equation of the
characteristic function, it is
\begin{eqnarray}
\frac{\partial \chi }{\partial t} &=&-\sum_{jk}(\eta _{jk}\mu _j^{*}\frac{%
\partial \chi }{\partial \mu _k}+\eta _{jk}^{*}\mu _j\frac{\partial \chi }{%
\partial \mu _k^{*}})  \label{wave} \\
&&-\frac 12\sum_j\Gamma _j\{\left| \mu _j\right| \frac{\partial \chi }{%
\partial \left| \mu _j\right| }+((2\overline{n}_j+1)\left| \mu _j\right|
^2-w_j^{*}\mu _j^2-w_j\mu _j^{*2})\chi \}  \nonumber \\
&&+\frac 12\sum_j\gamma _j\frac{\partial ^2\chi }{\partial \theta _j^2}.
\nonumber
\end{eqnarray}
Where we denote $\mu _j$ as $\left| \mu _j\right| e^{i\theta _j},$ and we
should take care about that the variables are $\mu _j$ and $\mu _j^{*}$ in
the amplification while they are $\left| \mu _j\right| $ $,\theta _j$ in the
damping.

\section{Solutions of some special cases}

Firstly, let us consider $\Gamma _j=\gamma _j=0$ for all $j,$ the solution
to the amplification is
\begin{equation}
\chi (\mu ,\mu ^{*},t)=\chi (\nu ,\nu ^{*},0),
\end{equation}
with
\begin{equation}
(\nu ,\nu ^{*})=(\mu ,\mu ^{*})\left(
\begin{array}{ll}
\cosh ^{*}(\left| \eta \right| t) & -\eta ^{*}\frac{\sinh (\left| \eta
\right| t)}{\left| \eta \right| } \\
-\frac{\sinh (\left| \eta \right| t)}{\left| \eta \right| }\eta & \cosh
(\left| \eta \right| t)
\end{array}
\right) ,
\end{equation}
where the matrix cosh and sinh functions are defined as\cite{Corney}
\begin{eqnarray}
\cosh \left| \xi \right| &=&I+\frac 1{2!}\xi \xi ^{*}+\frac 1{4!}(\xi \xi
^{*})^2+\cdots , \\
\frac{\sinh \left| \xi \right| }{\left| \xi \right| }\xi &=&\xi +\frac
1{3!}\xi \xi ^{*}\xi +\frac 1{5!}(\xi \xi ^{*})^2\xi +\cdots .  \nonumber
\end{eqnarray}

Then suppose $\eta =0$ and $\gamma _j=0$ for all $j,$ the solution to the
amplitude damping equation of $\chi $ is
\begin{equation}
\chi (\mu ,t)=\chi (\mu e^{-\frac{\Gamma t}2},0)\exp \{-\sum_j(1-e^{-\Gamma
_jt})[(\overline{n}_j+\frac 12)\left| \mu _j\right| ^2-\frac 12w_j^{*}\mu
_j^2-\frac 12w_j\mu _j^{*2}]\}.
\end{equation}
Where $\mu e^{-\frac{\Gamma t}2}$ is the abbreviation of $(\mu _1e^{-\frac{%
\Gamma _1t}2},\mu _2e^{-\frac{\Gamma _2t}2},\cdots ,\mu _se^{-\frac{\Gamma
_st}2})$. Next, suppose $\eta =0$ and $\Gamma _j=0$ for all $j,$ the
solution to the phase equation of $\chi $ then will be
\begin{equation}
\chi (\mu ,\mu ^{*},t)=\int dx\chi (\mu e^{ix},\mu
^{*}e^{-ix},0)\prod_i(2\pi \gamma _it)^{-1/2}\exp (-\frac{x_i^2}{2\gamma _it}%
),
\end{equation}
where $\mu e^{ix}$ is the abbreviation of $(\mu _1e^{ix_1},\mu
_2e^{ix_2},\cdots ,\mu _se^{ix_s})$. The simultaneous amplitude and phase
damping ($\eta =0$) for any initial characteristic function is (for $w_j=0$)
\cite{Chen}
\begin{eqnarray}
\chi (\mu ,\mu ^{*},t) &=&\int dx\chi (\mu e^{-\frac{\Gamma t}2+ix},\mu
^{*}e^{-\frac{\Gamma t}2-ix},0)\prod_j(2\pi \gamma _jt)^{-1/2} \\
&&\exp [-\frac{x_j^2}{2\gamma _jt}-(1-e^{-\Gamma _it})(\overline{n}_j+\frac
12)\left| \mu _j\right| ^2].  \nonumber
\end{eqnarray}
The density matrix then can be obtained by making use of operator integral.

\section{The parametric amplifier and the amplitude damping}

The diffusion equation (\ref{wave}) now is with its $\gamma _j=0$ for all $j$%
. Suppose the solution to the diffusion equation is
\begin{equation}
\chi (\mu ,\mu ^{*},t)=\chi (\nu ,\nu ^{*},0)\exp [\frac 12(\nu ,-\nu
^{*})\left(
\begin{array}{ll}
\alpha & \beta ^{*} \\
\beta & \alpha ^{*}
\end{array}
\right) (\nu ^{*},-\nu )^T-\frac 12(\mu ,-\mu ^{*})\left(
\begin{array}{ll}
\alpha & \beta ^{*} \\
\beta & \alpha ^{*}
\end{array}
\right) (\mu ^{*},-\mu )^T],
\end{equation}
where $\nu =\mu M+\mu ^{*}N,$ with $M$ and $N$ being time varying matrices. $%
\alpha $ and $\beta $ are constant matrices and $\alpha ^{\dagger }=\alpha ,$
$\beta =\beta ^T$. $M$ and $N$ are the solutions of the following matrix
equations
\begin{eqnarray}
\frac{dM}{dt} &=&-\eta ^{*}N-\frac \Gamma 2M,  \label{we1} \\
\frac{dN}{dt} &=&-\eta M-\frac \Gamma 2N,  \label{we2}
\end{eqnarray}
where $\Gamma =diag\{\Gamma _1,\Gamma _2,\cdots ,\Gamma _s\}.$ While $\alpha
$ and $\beta $ are the solution of the following matrix equations
\begin{eqnarray}
2\eta \alpha +2\alpha ^{*}\eta -\Gamma \beta -\beta \Gamma +\Gamma w+w\Gamma
&=&0,  \label{we3} \\
\Gamma \alpha +\alpha \Gamma -2\eta ^{*}\beta -2\beta ^{*}\eta -\Gamma (%
\overline{n}+\frac 12)-(\overline{n}+\frac 12)\Gamma &=&0.  \label{we4}
\end{eqnarray}
where $w$ $=diag\{w_1,w_2,\cdots ,w_s\},\overline{n}=diag\{\overline{n}_1,%
\overline{n}_2,\cdots ,\overline{n}_s\}.$ The constant matrices $\alpha $
and $\beta $ can be worked out as the solution of linear algebraic equations
(\ref{we3}) and (\ref{we4}). What left is to solve matrix equations (\ref
{we1}) and (\ref{we2}). There are two situations that the equations are
solvable. The first case is that all the modes undergo the same amplitude
damping, that is $\Gamma _1=\Gamma _2=\cdots =\Gamma _s$, thus $\Gamma
=\Gamma _1\mathbf{I}_s.$ $\Gamma $ commutes with any matrix. Equations (\ref
{we1}) and (\ref{we2}) have solution
\begin{eqnarray}
M &=&e^{-\frac 12\Gamma t}\cosh ^{*}(\left| \eta \right| t),  \label{we5} \\
N &=&-e^{-\frac 12\Gamma t}\frac{\sinh (\left| \eta \right| t)}{\left| \eta
\right| }\eta .  \label{we6}
\end{eqnarray}
The solution to the one-mode situation is simple. In the two-mode situation,
$M$ and $N$ can be further simplified. For $\eta $ is a symmetric matrix, we
can express it as $\eta =\eta _0\sigma _0+\eta _1\sigma _1+\eta _3\sigma _3,$
where $\sigma _0=\mathbf{I}_2,$ $\sigma _i$ $(i=1,2,3)$ are Pauli matrices.
The $\sigma _2$ term is nulled by the symmetry of $\eta .$ By using of the
algebra of Pauli matrices, we arrive at the results:
\begin{eqnarray}
\cosh (\left| \eta \right| t) &=&\sigma _0\cosh (\sqrt{(C+A)/2}t)\cosh (%
\sqrt{(C-A)/2}t)  \label{we7} \\
&&+\sinh (\sqrt{(C+A)/2}t)\sinh (\sqrt{(C-A)/2}t)\overrightarrow{\sigma }%
\cdot \overrightarrow{b},  \nonumber \\
\frac{\sinh (\left| \eta \right| t)}{\left| \eta \right| }\eta &=&\frac
12\sigma _0[\sinh (\sqrt{(C+B)}t)/\sqrt{(C+B)}+\sinh (\sqrt{(C-B)}t)/\sqrt{%
(C-B)}]\eta  \label{we8} \\
&&+\frac 12[\sinh (\sqrt{(C+B)}t)/\sqrt{(C+B)}-\sinh (\sqrt{(C-B)}t)/\sqrt{%
(C-B)}]\overrightarrow{\sigma }\cdot \overrightarrow{b}\eta .  \nonumber
\end{eqnarray}
Where $C=\left| \eta _0\right| ^2+\left| \eta _1\right| ^2+\left| \eta
_3\right| ^2,$ $A=\left| \eta _0^2-\eta _1^2-\eta _3^2\right| ,$ $B=\sqrt{%
C^2-A^2}.$ $\overrightarrow{b}$ is a unit vector and it is equal to $(\eta
_0\eta _1^{*}+\eta _1\eta _0^{*},i\eta _3\eta _1^{*}-i\eta _1\eta
_3^{*},\eta _0\eta _3^{*}+\eta _3\eta _0^{*})/B.$ Thus the characteristic
function is simplified. If the initial state is Gaussian. The correlation
matrix of the time dependent state can be obtained explicitly.

The second case is that $\eta $ is a real matrix while the amplitude damping
can be different for each mode. The solution of equations (\ref{we1}) and (%
\ref{we2}) will be
\begin{eqnarray}
M &=&\frac 12[\exp (-\eta t-\frac{\Gamma t}2)+\exp (\eta t-\frac{\Gamma t}%
2)], \\
N &=&\frac 12[\exp (-\eta t-\frac{\Gamma t}2)-\exp (\eta t-\frac{\Gamma t}%
2)].
\end{eqnarray}
For two-mode situation, this $M$ and $N$ can be simplified to
\begin{eqnarray}
M &=&\frac 12e^{-C_1t}[\cosh (B_1t)\sigma _0-\sinh (B_1t)\overrightarrow{%
\sigma }\cdot \overrightarrow{b_1}] \\
&&+\frac 12e^{-C_2t}[\cosh (B_2t)\sigma _0-\sinh (B_2t)\overrightarrow{%
\sigma }\cdot \overrightarrow{b_2}],  \nonumber \\
N &=&\frac 12e^{-C_1t}[\cosh (B_1t)\sigma _0-\sinh (B_1t)\overrightarrow{%
\sigma }\cdot \overrightarrow{b_1}] \\
&&-\frac 12e^{-C_2t}[\cosh (B_2t)\sigma _0-\sinh (B_2t)\overrightarrow{%
\sigma }\cdot \overrightarrow{b_2}].  \nonumber
\end{eqnarray}
where $C_{1,2}=\pm \eta _0+\frac 14(\Gamma _1+\Gamma _2);$ $B_{1,2}=\sqrt{%
\eta _1^2+(\frac 14(\Gamma _1-\Gamma _2)\pm \eta _3)^2};$ $\overrightarrow{b}%
_{1,2}=(\pm \eta _1,0,\pm \eta _3+\frac 14(\Gamma _1-\Gamma _2))/B_{1,2}.$

\section{One mode Gaussian system}

In this and next sections, we will apply the solutions of previous section
to Gaussian states. In the case of the parametric amplifier and the
amplitude damping, we consider the case of $\Gamma _1=\Gamma _2=\cdots
=\Gamma _s=\Gamma $ (hereafter $\Gamma $ is a number, not the matrix used in
previous sections). If the initial state is Gaussian, its characteristic
function has the form of $\chi (\mu ,\mu ^{*},0)=\exp [\mu m^{\dagger
}(0)-\mu ^{*}m^T(0)-\frac 12(\mu ,-\mu ^{*})\gamma (0)(\mu ^{*},-\mu )^T],$
the state will keep to be a Gaussian state in later evolution. The complex
correlation matrix (CM) $\gamma (0)$ should be chosen in a fashion that the
intial state is physical. The time evolution of the complex CM is
\begin{equation}
\gamma (t)=e^{-\Gamma t}\left[
\begin{array}{ll}
\cosh ^{*}(\left| \eta \right| t) & \eta ^{*}\frac{\sinh (\left| \eta
\right| t)}{\left| \eta \right| } \\
\frac{\sinh (\left| \eta \right| t)}{\left| \eta \right| }\eta & \cosh
(\left| \eta \right| t)
\end{array}
\right] (\gamma (0)-\left[
\begin{array}{ll}
\alpha & \beta ^{*} \\
\beta & \alpha ^{*}
\end{array}
\right] )\left[
\begin{array}{ll}
\cosh ^{*}(\left| \eta \right| t) & \eta ^{*}\frac{\sinh (\left| \eta
\right| t)}{\left| \eta \right| } \\
\frac{\sinh (\left| \eta \right| t)}{\left| \eta \right| }\eta & \cosh
(\left| \eta \right| t)
\end{array}
\right] +\left[
\begin{array}{ll}
\alpha & \beta ^{*} \\
\beta & \alpha ^{*}
\end{array}
\right] .
\end{equation}
The time evolution of the complex first moment $m$ is
\begin{equation}
m(t)=e^{-\frac 12\Gamma t}[m(0)\cosh ^{*}(\left| \eta \right| t)+m^{*}(0)%
\frac{\sinh (\left| \eta \right| t)}{\left| \eta \right| }\eta ].
\end{equation}

The solution of the characteristic function for one mode system is
characterized by Eq. (\ref{we5}) and Eq. (\ref{we6}) where $\eta $ is simply
a complex number now, and (for $\Gamma \neq 2\left| \eta \right| $)
\begin{eqnarray}
\alpha &=&\alpha ^{*}=\frac \Gamma {\Gamma ^2-4\left| \eta \right|
^2}[\Gamma (\overline{n}+\frac 12)+\eta ^{*}w+\eta w^{*}], \\
\beta &=&\frac 1{\Gamma ^2-4\left| \eta \right| ^2}[2\Gamma \eta (\overline{n%
}+\frac 12)+(\Gamma ^2-2\left| \eta \right| ^2)w+2\eta ^2w^{*}].
\end{eqnarray}
The result reduces to that of \cite{Corney} when $w$ $=0,\overline{n}=0$ and
$\eta $ is real. The degree of the mixedness of a quantum state $\rho $ is
characterized by means of the so called purity $\mu _p=Tr\rho ^2$. With
characteristic function, we have $\mu _p=\int \frac{d^2\mu }\pi $ $\left|
\chi (\mu ,\mu ^{*})\right| ^2$. For one mode Gaussian state, it reads
\begin{equation}
\mu _p(t)=\frac 1{2\sqrt{(\text{Re}\gamma _1{}(t))^2-\left| \gamma
_2(t)\right| ^2}},\text{ with }\gamma (t)=\left[
\begin{array}{ll}
\gamma _1(t) & \gamma _2^{*}(t) \\
\gamma _2(t) & \gamma _1^{*}(t)
\end{array}
\right] .
\end{equation}
We now consider the evolution of coherent state, squeezed state and thermal
state. The coherent state is given by $\exp [m(0)a^{\dagger
}-m^{*}(0)a]\left| 0\right\rangle ,$the complex CM is $\gamma
(0)=diag\{\frac 12,\frac 12\}.$ The squeezed state is given by $\exp [(\zeta
a^{\dagger 2}-\zeta ^{*}a^2)/2]\left| 0\right\rangle ,$ where $\zeta =r\exp
[i\varphi ],$the complex CM is $\gamma (0)=\frac 12[\cosh (2r)\sigma
_0+\sinh (2r)\cos \varphi \sigma _1+\sinh (2r)\sin \varphi \sigma _2].$ The
thermal state is given by $\sum_{n=0}^\infty [N/(N+1)]^n\left|
n\right\rangle \left\langle n\right| $ with $N$ is the average photon
number, the complex CM is $\gamma (0)=diag\{N+\frac 12,N+\frac 12\}.$

For $\Gamma >2\left| \eta \right| ,$ all these states (and other Gaussian
initial states) will tend to a Gaussian state with complex CM $\gamma
(\infty )=\left[
\begin{array}{ll}
\alpha & \beta ^{*} \\
\beta & \alpha
\end{array}
\right] $ after a sufficient large evolution time. The ultimate purity is
\begin{equation}
\mu _p(\infty )=\frac 12\{\frac 1{\Gamma ^2-4\left| \eta \right| ^2}[\Gamma
^2(\overline{n}+\frac 12)^2-\left| w\right| ^2(\Gamma ^2-2\left| \eta
\right| ^2)-\eta ^{*2}w^2-\eta ^2w^{*2}\}^{-\frac 12}.
\end{equation}
When the phase angle of the amplification $\eta $ is equal to the phase
angle of the 'squeezed' environment $w$ , the ultimate purity is maximized,
which is
\begin{equation}
\mu _{p\max }(\infty )=\frac 12\{\frac{\Gamma ^2}{\Gamma ^2-4\left| \eta
\right| ^2}(\overline{n}+\frac 12)^2-\left| w\right| ^2\}^{-\frac 12}.
\end{equation}
The ultimate purity is a monotonically decreasing function of the
amplification $\left| \eta \right| $.

For $\Gamma <2\left| \eta \right| ,$ we consider the case of $w$ $=0$ and $%
\eta $ is real and positive for simplicity. When the initial state is a
thermal state with $\gamma (0)=diag\{N+\frac 12,N+\frac 12\}$ (coherent
state is the special case of $N=0$ when CM is concerned). Denote $\cosh
r_0=2\eta /\sqrt{4\eta ^2-\Gamma ^2}$, then
\begin{eqnarray}
\gamma (t) &=&e^{-\Gamma t}(N+\frac 12)[\cosh (2\eta t)\sigma _0+\sinh
(2\eta t)\sigma _1]  \nonumber \\
&&+(\overline{n}+\frac 12)\sinh r_0[e^{-\Gamma t}(\sinh (2\eta t+r_0)\sigma
_0+\cosh (2\eta t+r_0)\sigma _1)-(\sinh r_0\sigma _0+\cosh r_0\sigma _1)].
\end{eqnarray}
The state is a squeezed thermal state, with its purity
\begin{equation}
\mu _p(t)=\frac 12\{[N^{\prime }e^{(2\eta -\Gamma )t}+\overline{n}^{\prime
}e^{r_0}(e^{(2\eta -\Gamma )t}-1)][N^{\prime }e^{-(2\eta +\Gamma )t}+%
\overline{n}^{\prime }e^{-r_0}(1-e^{-(2\eta +\Gamma )t})]\}^{-\frac 12},
\end{equation}
where $N^{\prime }=N+\frac 12,$ $\overline{n}^{\prime }=(\overline{n}+\frac
12)\sinh r_0.$ For large $t,$ $\mu _p(t)\approx \frac 12\exp [-(\eta -\Gamma
/2)t]/\sqrt{\overline{n}^{\prime 2}+\overline{n}^{\prime }N^{\prime }e^{-r_0}%
}.$ The purity is an decreasing function of the amplification $\eta $ at
large $t.$ When the initial state is a squeezed state with squeezed
parameter $\zeta =r\exp [i\varphi ],$ we consider the simple case of $%
\varphi =0,$then
\begin{equation}
\mu _p(t)=\frac 12\{[\frac 12e^{(2\eta -\Gamma )t+2r}+\overline{n}^{\prime
}e^{r_0}(e^{(2\eta -\Gamma )t}-1)][\frac 12e^{-(2\eta +\Gamma )t-2r}+%
\overline{n}^{\prime }e^{-r_0}(1-e^{-(2\eta +\Gamma )t})]\}^{-\frac 12}.
\end{equation}
For large $t,$ $\mu _p(t)\approx \frac 12\exp [-(\eta -\Gamma /2)t]/\sqrt{%
\overline{n}^{\prime 2}+\overline{n}^{\prime }e^{2r-r_0}/2}.$

\section{Two mode Gaussian system}

The algebra equation of $\alpha $ and $\beta $ in two mode system is
complicated in general situation. To investigate the entanglement property
of the amplifier, we will set $\eta _0=\eta _3=0$ which corresponds to no
single mode amplification. Thus $\eta =\eta _1\sigma _1$, it corresponds to
two mode amplification. The solution is simplified to (with $\Gamma
_1=\Gamma _2=\Gamma $, and the two modes undergo the same noise ($\overline{n%
}=\overline{n}_0\mathbf{I}_2$) for simplicity)
\begin{eqnarray}
\alpha &=&\frac \Gamma {\Gamma ^2-4\left| \eta _1\right| ^2}\left[
\begin{array}{ll}
\Gamma (\overline{n}_0+\frac 12), & \eta _1w_b+\eta _1^{*}w_a^{*} \\
\eta _1w_a+\eta _1^{*}w_b^{*} & \Gamma (\overline{n}_0+\frac 12)
\end{array}
\right] , \\
\beta &=&\frac 1{\Gamma (\Gamma ^2-4\left| \eta _1\right| ^2)}\left[
\begin{array}{ll}
(\Gamma ^2-2\left| \eta _1\right| ^2)w_a+\eta _1^2w_b^{*} & 2\eta _1\Gamma (%
\overline{n}_0+\frac 12) \\
2\eta _1\Gamma (\overline{n}_0+\frac 12) & (\Gamma ^2-2\left| \eta _1\right|
^2)w_b+\eta _1^2w_a^{*}
\end{array}
\right] ,
\end{eqnarray}
where we have denoted $w$ $=diag\{w_a,w_b\}$ for the two modes $\ a$ and $b.$
If the initial state is Gaussian, the state will keep to be a Gaussian state
in later evolution.

For $\Gamma >2\left| \eta _1\right| $, the state will tend to a Gaussian
state which is characterized by the residue complex CM $\gamma (\infty )$
after a sufficient larger evolution time. The Peres-Horodecki criterion \cite
{Simon} \cite{Duan} for separability of the state is an inequality on the
real parameter CM $\gamma _{re}(\infty ),$ which is
\begin{equation}
\gamma _{re}(\infty )=\left[
\begin{array}{ll}
\gamma _a & \gamma _c \\
\gamma _c^T & \gamma _b
\end{array}
\right] =L\left[
\begin{array}{ll}
\alpha & \beta ^{*} \\
\beta & \alpha ^{*}
\end{array}
\right] L^{\dagger },
\end{equation}
with
\begin{equation}
L=\frac 1{\sqrt{2}}\left[
\begin{array}{llll}
i & 0 & i & 0 \\
1 & 0 & -1 & 0 \\
0 & i & 0 & i \\
0 & 1 & 0 & -1
\end{array}
\right] .
\end{equation}
Thus we have $\gamma _a=\alpha _a\sigma _0-$Im$\beta _a\sigma _1+$Re$\beta
_a\sigma _3,$ $\gamma _b=\alpha _b\sigma _0-$Im$\beta _b\sigma _1+$Re$\beta
_b\sigma _3,$ $\gamma _c=$Re$\alpha _c\sigma _0-$Im$\beta _c\sigma _1-i$Im$%
\alpha _c\sigma _2+$Re$\beta _c\sigma _3,$ where we have denoted
\begin{equation}
\alpha =\left[
\begin{array}{ll}
\alpha _a & \alpha _c \\
\alpha _c^{*} & \alpha _b
\end{array}
\right] ,\text{ }\beta =\left[
\begin{array}{ll}
\beta _a & \beta _c \\
\beta _c & \beta _b
\end{array}
\right] .
\end{equation}
The separable criterion takes the form of $\det \gamma _a\det \gamma
_b+(\frac 14$ -$\left| \det \gamma _c\right| )^2$-$tr(\gamma _aJ\gamma
_cJ\gamma _bJ\gamma _c^TJ)\geq $ $\frac 14(\det \gamma _a+\det \gamma _b)$
\cite{Simon}, with $J=i\sigma _2.$ Hence in the form of $\alpha $ and $\beta
$, it will be
\begin{equation}
\det \gamma _a^{\prime }\det \gamma _b^{\prime }+(\frac 14-\left| \det
\gamma _c^{\prime }\right| )^2-tr(\gamma _a^{\prime }\sigma _3\gamma
_c^{\prime }\sigma _3\gamma _b^{\prime }\sigma _3\gamma _c^{\prime \dagger
}\sigma _3)\geq \frac 14(\det \gamma _a^{\prime }+\det \gamma _b^{\prime }),
\end{equation}
where
\begin{equation}
\gamma _i^{\prime }=\left[
\begin{array}{ll}
\alpha _i & \beta _i^{*} \\
\beta _i & \alpha _i^{*}
\end{array}
\right] ,\text{ }i=a,b,c.
\end{equation}
We now consider the situation of $w$ $=0$ for simplicity. Then $\alpha
_a=\alpha _b=$ $\frac{\Gamma ^2(\overline{n}_0+\frac 12)}{\Gamma ^2-4\left|
\eta _1\right| ^2},\alpha _c=0;$ $\beta _c=\frac{2\eta _1^{*}\Gamma (%
\overline{n}_0+\frac 12)}{\Gamma ^2-4\left| \eta _1\right| ^2},$ $\beta
_a=\beta _b=0,$ the CM $\gamma _{re}(\infty )$ can be transformed to the
standard form \cite{Duan} by local rotation. The standard form CM\ is $%
\gamma _{re}^s(\infty )=\frac{\Gamma (\overline{n}_0+\frac 12)}{\Gamma
^2-4\left| \eta _1\right| ^2}(\Gamma \sigma _0\otimes \sigma _0+2\left| \eta
_1\right| \sigma _1\otimes \sigma _3).$The inseparability criterion reads $%
\alpha _a-\left| \beta _c\right| <\frac 12$, which is
\begin{equation}
\Gamma \overline{n}_0<\left| \eta _1\right| <\Gamma /2.  \label{we9}
\end{equation}
The possible entanglement appears only when $\overline{n}_0<\frac 12$ . The
entanglement of formation (EoF) of the inseparable state will be \cite
{Giedke} \cite{Chen1}
\begin{equation}
E_f=g(\Delta (2\alpha _a-2\left| \beta _c\right| ))=g(\Delta (\frac{\Gamma (2%
\overline{n}_0+1)}{\Gamma +2\left| \eta _1\right| })),  \label{we10}
\end{equation}
where $\Delta (z)=(z+z^{-1}-2)/4$ and $g(x)=(x+1)\log _2(x+1)-x\log _2x$ is
the bosonic entropy function. $g(x)$ is a monotonically increasing function
of $x,$ thus the entanglement is a monotonically increasing function of the
amplification $\left| \eta _1\right| $( note that $\alpha _a-\left| \beta
_c\right| <\frac 12$). The entanglement tends to its supremum when $%
\overline{n}_0=0$ and $\left| \eta _1\right| \rightarrow \Gamma /2.$ The
supremum is $E_f^{\sup }=g(\frac 18)=0.5662$ .

For $\Gamma <2\left| \eta _1\right| $, we consider case of $w$ $=0$ and $%
\eta _1$ is real and positive for simplicity. From equations (\ref{we7}) and
(\ref{we8}), the time evolution solution of complex CM will be \
\begin{eqnarray}
\gamma (t) &=&e^{-\Gamma t}[\cosh (\eta _1t)\sigma _0\otimes \sigma _0+\sinh
(\eta _1t)\sigma _1\otimes \sigma _1][\gamma (0)-(\alpha _a\sigma _0\otimes
\sigma _0+\beta _c\sigma _1\otimes \sigma _1)]  \nonumber \\
&&[\cosh (\eta _1t)\sigma _0\otimes \sigma _0+\sinh (\eta _1t)\sigma
_1\otimes \sigma _1]+\alpha _a\sigma _0\otimes \sigma _0+\beta _c\sigma
_1\otimes \sigma _1.
\end{eqnarray}
Where $\beta _c$ is real and positive. If the initial state is a two mode
squeezed thermal state with real CM $\gamma _{re}(0)=(N+\frac 12)(\cosh
2r\sigma _0\otimes \sigma _0+\sinh 2r\sigma _1\otimes \sigma _3),$ then $%
\gamma _{re}(t)=\gamma _{re1}(t)\sigma _0\otimes \sigma _0+$ $\gamma
_{re2}(t)\sigma _1\otimes \sigma _3,$ with
\begin{eqnarray}
\gamma _{re1}(t) &=&e^{-\Gamma t}[(N+\frac 12)\cosh 2(\eta _1t+r)-\alpha
_a\cosh (2\eta _1t)-\beta _c\sinh (2\eta _1t)]+\alpha _a \\
\gamma _{re2}(t) &=&e^{-\Gamma t}[(N+\frac 12)\sinh 2(\eta _1t+r)-\alpha
_a\sinh (2\eta _1t)-\beta _c\cosh (2\eta _1t)]+\beta _c
\end{eqnarray}
The inseparable criterion for the state is $\gamma _{re1}(t)-\gamma
_{re2}(t)<\frac 12,$ that is
\begin{equation}
(N+\frac 12)\exp [-(2\eta _1+\Gamma )t-2r]+(\alpha _a-\beta _c)\{1-\exp
[-(2\eta _1+\Gamma )t]\}<\frac 12.  \label{we11}
\end{equation}
The EoF of the inseparable state will be
\begin{equation}
E_f(t)=g(\Delta (2\gamma _{re1}(t)-2\gamma _{re2}(t))).
\end{equation}
\begin{figure}[tbp]
\includegraphics[width=2.5in]{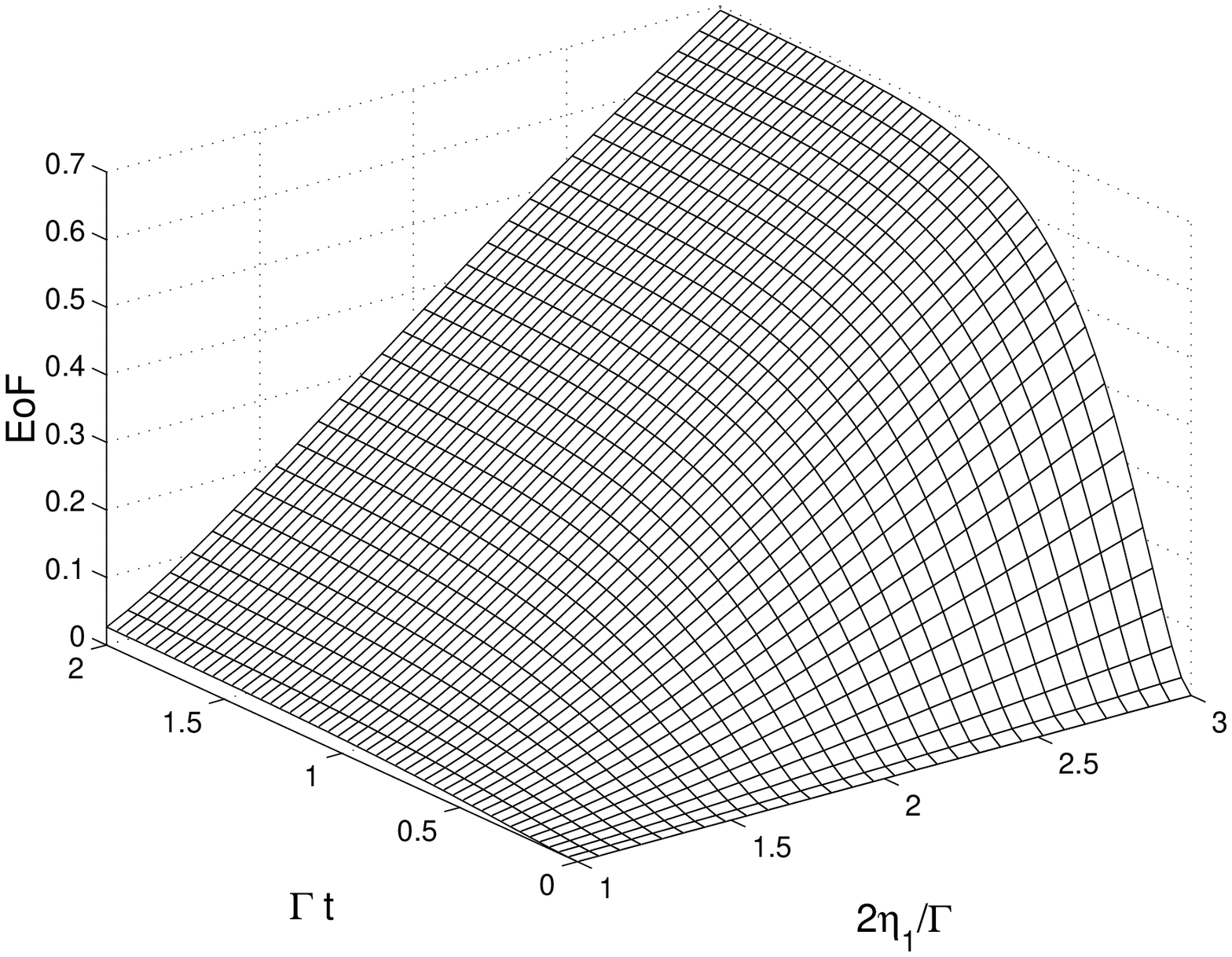}
\caption{Entanglement of formation of the two mode amplification and
damping, with the environment noise $\overline{n}_0=0.4$ . The initial state
is vacuum state. }
\end{figure}
\begin{figure}[tbp]
\includegraphics[width=2.5in]{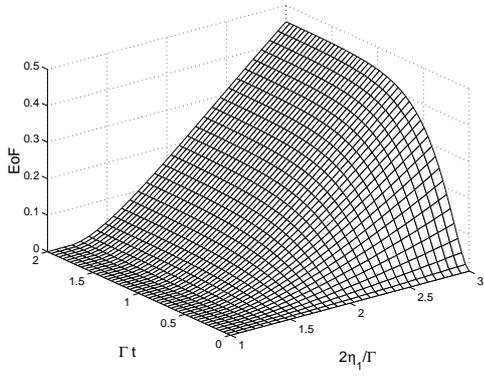}
\caption{Entanglement of formation of the two mode amplification and
damping, with the environment noise $\overline{n}_0=0.6$ . The initial state
is vacuum state. }
\end{figure}
When $t\rightarrow \infty ,$ EoF saturates at
\begin{equation}
E_f(\infty )=g(\Delta (\frac \Gamma {\Gamma +2\eta _1}[2\overline{n}_0+1])).
\end{equation}
If the initial state is vacuum state, then $N=0,r=0.$ Inequality (\ref{we11}%
) reduces to $\alpha _a-\beta _c<\frac 12$ $.$ It can be further simplified
to
\begin{equation}
\eta _1>\max \{\overline{n}_0\Gamma ,\frac \Gamma 2\}.  \label{we12}
\end{equation}
EoFs for $\overline{n}_0=0.4,0.6$ are shown in Fig.1 and Fig.2. We can see
that EoF is a monotonically increasing function of the amplification $\eta
_1,$ $E_f(\infty )$ has the same expression for $\eta _1>$ $\frac \Gamma 2$
and $\eta _1<$ $\frac \Gamma 2$ (see (\ref{we10}))$.$ $E_f(\infty )$ is
shown in Fig.3 (It is for any Gaussian initial state, not only for vacuum
initial state). The inseparable criterion inequalities (\ref{we9}) and (\ref
{we12}) can be combine
\begin{equation}
\eta _1/\Gamma >\overline{n}_0.
\end{equation}
\begin{figure}[tbp]
\includegraphics[width=2.5in]{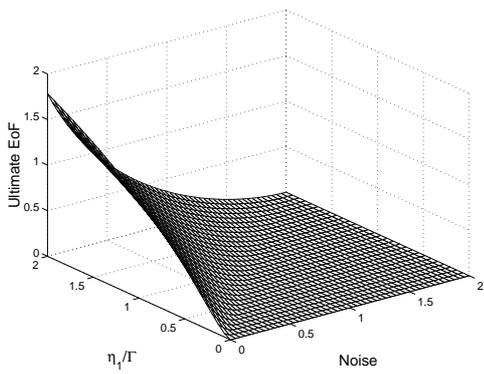}
\caption{The ultimate entanglement of formation with respect to
amplification damping ratio $\eta _1/\Gamma $ and environment noise $%
\overline{n}_0$}
\end{figure}

\section{Conclusion}

The master equation of quantum continuous variable system is converted to
the equation of quantum characteristic function. It turn out to be a linear
partial differential equation about the characteristic function. The time
evolution solution can be obtained exactly for any initial quantum optical
state in several cases. The solvable cases include (1) parametric
amplification \cite{Corney} , (2) amplitude damping with thermal or squeezed
noise \cite{Serafini} , (3) simultaneously amplitude and phase damping
together with thermal noise\cite{Chen}, (4) simultaneous multi-mode
parametric amplification and amplitude damping with thermal or squeezed
noise when each mode undergoes the same strength of damping, (5)
simultaneous multi-mode real parametric amplification and amplitude damping
with thermal or squeezed noise.

The applications to one mode and two mode Gaussian initial conditions are
investigated. In one mode Gaussian system, the purity of the evolution state
is monotonically decreasing with the amplification. In the situation of less
amplification ($\left| \eta \right| <\Gamma /2$), the ultimate purity is
maximized when the phase of the amplification $\eta $ matches with the phase
of the 'squeezed' environment $w$ $.$ In two mode Gaussian system, the
entanglement of formation monotonically increases with the two mode
amplification. In the situation of less amplification ($\left| \eta
_1\right| <\Gamma /2$), the supremum of the entanglement of formation is
given. In the situation of over amplification ($\left| \eta _1\right|
>\Gamma /2$), after a sufficient large time, the entanglement of formation
saturates. The ultimate entanglement of formation is given as a function of
the amplification damping ratio and the noise. It is independent of the
initial Gaussian state. The ultimate state is separable when the
amplification damping ratio is greater than the noise.

In real experiment, the parametric amplification and the damping may occur
successively. Thus the concatenate of above solutions will represent most of
the optical evolution system.

\section*{Acknowledgment}

Funding by the National Natural Science Foundation of China (under Grant No.
10575092, 10347119), Zhejiang Province Natural Science Foundation (under
Grant No. RC104265) and AQSIQ of China (under Grant No. 2004QK38) are
gratefully acknowledged.

\end{document}